\documentclass[aps,prl,twocolumn,showpacs]{revtex4}
\usepackage{amsmath}
\usepackage{epsfig}
\usepackage{amssymb}
\usepackage{dcolumn}
\usepackage{graphicx}
\usepackage{dcolumn}

\begin{document}
\date{\today}

\title{Localized nonlinear waves in systems with time- and space-modulated nonlinearities}

\author{Juan Belmonte-Beitia$^1$}

\author{V\'{\i}ctor  M. P\'erez-Garc\'{\i}a$^1$}

\author{Vadym Vekslerchik$^1$}

\author{Vladimir V. Konotop$^2$}

\affiliation{$^1$Departamento de Matem\'aticas, Escuela T\'ecnica
Superior de Ingenieros Industriales, and Instituto de Matem\'atica Aplicada a la Ciencia y la Ingenier\'{\i}a (IMACI),
Universidad de Castilla-La Mancha, 13071 Ciudad Real, Spain \\ $^2$Centro de F\'{\i}sica Te\'{o}rica e Computacional, Universidade de Lisboa,
 Complexo Interdisciplinar, Avenida Professor Gama Pinto 2, Lisboa
1649-003, Portugal and Departamento de F\'{\i}sica, Faculdade de Ci\^encias,
Universidade de Lisboa, Campo Grande, Ed. C8, Piso 6, Lisboa
1749-016, Portugal.}

\begin{abstract}
Using similarity transformations we construct explicit nontrivial solutions of nonlinear Schr\"odinger equations with potentials and nonlinearities depending on time and on the spatial coordinates. We present the general theory and use it to calculate explicitly non-trivial solutions such as periodic (breathers), resonant or quasiperiodically oscillating solitons. Some implications to the field of matter-waves are also discussed. 
\end{abstract}


\maketitle

\emph{Introduction}.- The concept of solitons was introduced in the seminal paper by Zabusky and Kruskal \cite{SoliGen}  to characterize  nonlinear solitary waves, which emerge from the delicate balance between nonlinearity and dispersion, and  preserve their localized shapes and velocities during propagation and after collisions. It was soon realized that solitons are remarkably generic objects arising in different contexts and were observed in many experiments, for example in nonlinear optics starting with the pioneering works~\cite{HT}. Being solutions of integrable nonlinear wave equations, standard solitons appear usually in translationally invariant systems involving nonlinearities which depend only on the relevant fields and do not have explicit dependence on the space and time variables, because such dependences typically break the symmetries of the system. However, those extra dependences of the nonlinear interactions open many new practical possibilities for generation and control of solitons, what has stimulated their study in the last thirty years. It was already recognized 
from the very first papers~\cite{old} that specific dependencies of the equation coefficients on the spatial spatial coordinates and on time can preserve the integrability. Moreover this gave an origin to new concepts, like for example, non-autonomous solitons~\cite{Serkin} highly relevant in nonlinear fiber optics. During the last few years these studies have seen an enormous revival in the context of systems ruled by nonlinear Schr\"odinger (NLS) equations because of their applications in the mean field theory of Bose-Einstein condensates (BECs)~\cite{Pit,book} and in nonlinear optics~\cite{Yuri,transformaciones1,Serkin}.

In BEC applications,  the matter waves being the natural outcome of the mean-field description~\cite{Pit} have attracted a great deal of interest in the last years \cite{book}. The possibility of using  Feschbach resonances to control the nonlinearities (see e.g. \cite{FB1,FB2}) has lead to the proposal of many different nonlinear phenomena induced by the manipulation of the scattering length either in time  \cite{Kono1,Ueda,Boris} or in space \cite{VGP}. In nonlinear optics, recent developments~\cite{similariton} have led to the discovery of a new classes of waves such as  the so-called optical similariton which arise when the interaction of nonlinearity, dispersion and gain in a high-power fibre amplifier causes the shape of an arbitrary input pulse to converge asymptotically to a pulse whose shape is self-similar. 

In this letter, we go beyond  previous studies considering space and time dependent nonlinearities and constructing different types of explicit   solutions including:  breathers, resonantly oscillating   and quasiperiodic
solitary waves. We provide explicit expressions for all those solutions, exploring 
realistic choices of nonlinearities which are experimentally feasible due to the flexible and precise control of the scattering length achievable in quasi-one-dimensional BECs~\cite{com} with tunable interactions.

\emph{General Theory.-} We consider physical systems ruled by the NLS equation 
\begin{equation}
  i \psi_{t} = - \psi_{xx} + v(t,x) \psi + g(t,x) \left| \psi \right|^{2} \psi,  
\label{eq-physical}
\end{equation}
 for an unknown complex function $\psi(t,x)$ with a potential $v(x,t)$ and nonlinearity  $g(x,t)$. We focus on spatially localized solutions for which  $\lim_{|x|\to\infty}\psi(t,x)=0$. Our goal is to reduce Eq. (\ref{eq-physical}) to the stationary NLS equation
\begin{equation}
  \mu \Phi = - \Phi_{XX} +G \left| \Phi \right|^{2} \Phi,
\label{eq-stationary}
\end{equation}
where $\Phi \equiv \Phi(X)$, $X \equiv X(t,x)$ is a function to be determined, $\mu$ is the eigenvalue of the nonlinear equation, and $G$ is a constant. In this letter, we explore the case $G=-1$, i.e., focusing or attractive nonlinearity. The case $G=1$ does not pose any new challenges and will be studied elsewhere. 

To connect solutions of Eq. (\ref{eq-physical}) with those of Eq. (\ref{eq-stationary}) we will use the transformation
\begin{equation}
  \psi(t,x) = \rho(t,x) e^{i\varphi (t,x)} \Phi\left( X(t,x) \right).
\label{psi-from-Phi}
\end{equation}
Other similarity transformations have been studied for NLS equations in different contexts \cite{VPK,transformaciones1,transformaciones2}. 

Requiring $\Phi(X)$ to satisfy  Eq. (\ref{eq-stationary}) 
and $\psi(t,x)$ to be a solution of Eq. (\ref{eq-physical}) with potential $v(t,x)$ and 
nonlinearity $g(t,x)$  we get the set of equations
\begin{eqnarray*}
 \rho \rho_{t} + \left( \rho^{2} \varphi_{x} \right)_{x}= 0, \,\,\, \left( \rho^{2} X_{x} \right)_{x}=0,\,\,\,  X_{t} + 2 \varphi_{x} X_{x}=0.
\label{syst}
\end{eqnarray*}
Let us take
\begin{equation}
\label{x_fin}
  X(t,x) = F(\xi),
 \qquad \text{with}\quad
  \xi(t,x) = \gamma(t) \, x + \delta(t),
\end{equation}
where $\gamma (t)$ is a positive definite function representing the inverse of the width of the localized solution, and $-\delta(t)/\gamma(t)$ is the position of the center of mass.  
After some algebra we obtain
\begin{equation}
\label{v_fin}
  v(t,x) =  
\frac{\rho_{xx}}{\rho}
  - \varphi_{t} - \varphi_{x}^{2} - \mu \frac{\gamma^4}{\rho^{4}},
\quad
  g(t,x)  = \frac{\gamma^{4}}{\rho^{6}},
\end{equation}
\begin{eqnarray}
\label{rho_fin}
  \rho(t,x) =  \sqrt{\frac{\gamma}{F'(\xi)} }, \quad
  \varphi(t,x)  =  
  -  \frac{\gamma_{t}} {4\gamma } x^{2} 
  -  \frac{\delta_{t}}{ 2\gamma} x+\alpha,
\end{eqnarray}
where $\alpha(t)$ is an arbitrary function of time.

Thus, choosing $\delta(t), \gamma(t)$ and $F(\xi)$ (or equivalently $\rho(x,t)$) we can generate pairs 
$g(x,t), v(x,t)$ for which the solutions of Eq. (\ref{eq-physical}) can be obtained from those of Eq. (\ref{eq-stationary}) using Eqs. (\ref{psi-from-Phi}). We will use this fact to construct soliton solutions with an interesting  nontrivial  behavior.

\begin{figure}
\epsfig{file=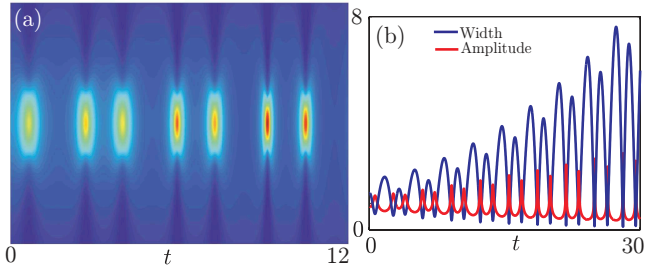, width=8.5cm}
 \caption{[Color online] Solution (\ref{resonant}), for $n=1$,   $\omega_0=2$, and $\varepsilon=0.5$. Initial data for Eq. (\ref{EP}) are $\nu(0)=\sqrt{2}$ and $\nu_{t}(0)=0$.  (a) Pseudocolor plot of $|\psi(x,t)|^2$  on the domain $x\in [-3,3]$ (b)  Width $\nu(t)$ (blue line) and amplitude $\sqrt{\gamma}$ (red line) {\it vs} $t$.
  \label{solucionesresonantescrec}}
\end{figure}

\emph{Localized nonlinearities.-}
As a first application of the method, we will study the case of localized nonlinearities. While  the choice of possible models is very rich, because of applications the possible applications of our results to BECs with optically controled  interactions \cite{FB2}, we will focus on the gaussian shaped nonlinearity  
\begin{equation}
\label{nolinealidad}
g(t,x)=-\gamma(t)\exp \left(-3\xi^{2}\right),
\end{equation}
and the potential including  a combination of harmonic and dipole traps, which are typical in BEC experiments
\begin{equation}
v(t,x)=\omega^2(t) x^{2}+f(t)x +h(t)-\mu\gamma^{2}(t)\exp(-2\xi^{2}), 
\end{equation}
where
\begin{subequations}
\begin{eqnarray}
\label{f1}
\omega^2(t)&=&\gamma^{4}+\left(\gamma_{tt}\gamma-2\gamma_{t}^{2}\right)/4\gamma^{2},\\
\label{f2}
f(t)&=&\left(4\gamma^{5}\delta+\delta_{tt}\gamma-2\delta_{t}\gamma_{t}\right)/2\gamma^{2},\\
\label{f3}
h(t)&=&\gamma^{2}(1+\delta^{2})-\delta_{t}^{2}/4\gamma^{2}-\alpha_{t}.
\end{eqnarray}
\end{subequations}

Our choice corresponds to 
$
\rho(t,x)=\sqrt{\gamma}\exp(\xi^{2}/2).
$

In the particular case where $\delta(t)\equiv 0$, $\mu=0$ and $\alpha(t)=\int\gamma^{2}(t)dt$, we obtain
\begin{eqnarray}
\label{potandnon}
v(t,x)=\omega^{2}(t)x^{2},\quad  
g(t,x)=-\gamma(t)e^{-3\gamma^{2}(t)x^{2}}.
\end{eqnarray}
Thus, we have a harmonic potential and a modulated in time gaussian nonlinearity which would correspond in BEC applications to a modulated laser beam controlling the interactions optically. In this paper we  focus on nonlinearities and potentials given by Eq. (\ref{potandnon}) but many other possibilities can be explored using the same ideas.

Defining $\nu=1/\gamma$ we rewrite  Eq. (\ref{f1}) in the form
\begin{equation}
\label{EP}
\nu_{tt}+4\omega^{2}(t)\nu= 4/\nu^{3},
\end{equation}
which is the Ermakov-Pinney equation \cite{theores1}. To discuss explicit solutions we choose  $\omega^{2}(t)$ of the form 
\begin{equation}
\label{omega}
\omega^{2}(t)=1+\varepsilon\cos(\omega_{0}t),
\end{equation}
 where $\varepsilon\in(-1,1)$ and $0\neq\omega_{0}\in\mathbb{R}$. After some algebra \cite{theores1}, we obtain from Eqs. (\ref{EP}) and (\ref{omega})  
  %
 \begin{eqnarray}
 \label{vgama}
 \gamma(t)= \left[2y_1^{2}(t)+2(y_2(t)/W)^{2}\right]^{-1/2}
 \end{eqnarray}
 %
where $y_{1,2}(t)$ are two linearly independent solutions of the Mathieu equation 
\begin{eqnarray}
\label{Mathieu}
y_{tt}+4\omega^2(t)y=0
\end{eqnarray}
and $W$ is their Wronskian. 

   \begin{figure}
   \epsfig{file=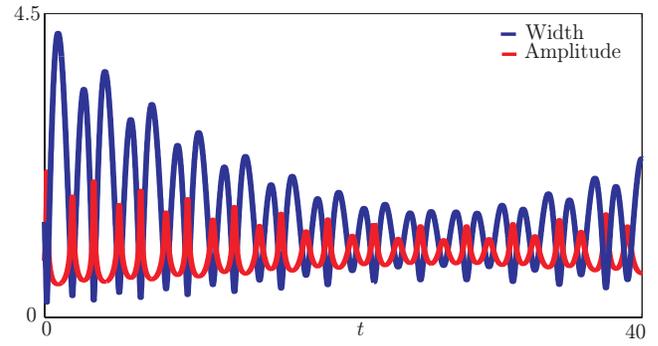,width=8.6cm}
 \caption{[Color online] Width (blue line) and amplitude (red line) versus time of the solution of Eq. (\ref{eq-physical}), given by Eq. (\ref{resonant}) for $n=1$, $\omega_0=2$, and $\varepsilon=0.5$. Initial data for Eq. (\ref{EP}) given by $\nu(0)=\sqrt{2}, \nu_{t}(0)=-6.125\sqrt{2}$. 
 \label{solucionesresonantesdec}}
\end{figure}

Thus it is Eq. (\ref{Mathieu}) that determines the temporal dynamics of the obtained solutions. Considering different choices of the parameters $\omega_0$ and $\varepsilon$, one can single out three different types of behaviors which can be classified as follows: (i) {\em periodic}, when $\varepsilon=0$ or in the frontiers bewteen the stability and instability regions of Eq. (\ref{Mathieu}), (ii) {\em resonant}, when $\varepsilon\neq 0$ and $y_{1,2}(t)$ belong to an instability region of Eq. (\ref{Mathieu}), and (iii) {\em quasi-periodic}, when $y_{1,2}(t)$ are in the stability region. Below we give examples of all these cases.

 \begin{figure}
  \epsfig{file=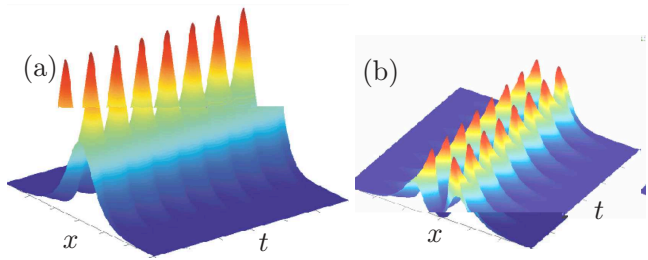,width=8.5cm}
 \caption{[Color online]  Plots of $|\psi(x,t)|^2$ for breathing solutions of Eq. (\ref{eq-physical}), given by Eqs. (\ref{resonant}), for $\varepsilon=0$ in Eq. (\ref{omega}), and $\gamma(t)$ given by Eq. (\ref{gamabreather}), corresponding to (a) $n=1$ (b) $ n=2$. For both cases, $x\in[-6,6]$, $t\in[0,10]$ and the initial data for Eq. (\ref{EP}) are $\nu(0)=\sqrt{2}, \nu_{t}(0)=0$. \label{soluciones}}
\end{figure}
Concerning the solutions of Eq. (\ref{eq-stationary}), since $\mu=0$, Eq. (\ref{eq-stationary}) becomes 
$   \Phi'' + \Phi^{3} = 0$, and its solution is given by
\begin{equation}
\Phi(X)= \eta \mathop{\mbox{sn}}(\eta X,k_{*})/2\mathop{\mbox{dn}}(\eta X,k_{*}),
\end{equation}
with arbitrary $\eta$ and $k_{*}=1/\sqrt{2}$.  We choose $X$ as
\begin{equation}
X(t,x)=\int_{-\infty}^{\gamma(t)x} e^{-\xi^{2}} d\xi ,
\end{equation}
which takes values in $(0,\sqrt{\pi})$. This choice of $X$ ensures vanishing of $\Phi$ when $x\rightarrow-\infty$. The imposition of zero boundary condition for $x\rightarrow+\infty$ leads to $\eta=2nK(k_{*})/\sqrt{\pi}$ where $K(k)$ is the elliptic integral. This leads to the family of solutions
\begin{equation}
\label{resonant}
\psi_{n} = n\sqrt{\gamma(t)}\exp\left(\frac 12 \gamma^{2}(t)x^{2}+i\varphi(t,x))\Phi_{1}(n\theta(t,x)\right) 
\end{equation}
where $n=1,2,...$,  $\varphi(t,x)=-(\gamma_{t}/4\gamma)x^2+\int\gamma^{2}(t)dt$,
\begin{equation}
\Phi_{1}(\theta)=\sqrt{ 2/\pi }K(k_{*}) \mathop{\mbox{sn}}(\theta,k_{*})/\mathop{\mbox{dn}}(\theta,k_{*}) 
\end{equation}
and $\theta(t,x)=\left(2K(k_{*})/\sqrt{\pi}\right)\int_{-\infty}^{\gamma(t)x}d\xi e^{-\xi^{2}}$. It is easy to see that $\psi(x,t)\rightarrow 0$ when $ |x|\rightarrow \infty$ in (\ref{resonant}), thus they correspond to localized solutions.  
 
\emph{Resonant solitons.-} The response of driven nonlinear systems to parametric perturbations is a field of interest which has recently also been investigated in the framework of the nonlinear dynamics of BECs 
\cite{theores1,theores2,expres}. 

Starting with the particular case $\omega_{0}=2$, in Eq. (\ref{omega}), we observe that for $\varepsilon \neq 0$ we are in the instability region, thus,  Eq. (\ref{resonant}) describes \emph{multisoliton resonant solutions}, since  $\psi_{n}(x,t)$ possesses $n-1$ zeros. 
In Fig. \ref{solucionesresonantescrec}(a), we plot a resonant solution corresponding to $n=1$, with $\omega_{0}=2$. Such resonant soliton has an increasing resonant behaviour. In Fig. \ref{solucionesresonantescrec}(b), we plot its width and amplitude versus time.  In Fig. \ref{solucionesresonantesdec}, we also plot the width and amplitude versus time for the same soliton but now with transitory decreasing resonant behaviour, for $n=1$.

It is remarkable that we are able to construct explicit resonant solitons 
in a very complicated scenario including a localized pulsating nonlinearity and a parametrically modulated trap along the longitudinal direction. Our solutions provide the first explicit resonant solution in a parametrically modulated one dimensional BEC.  A similar resonant behavior was observed in Ref.  \cite{expres} when driving a BEC, with positive scattering length, along the transverse direction.

  \begin{figure}
   \epsfig{file=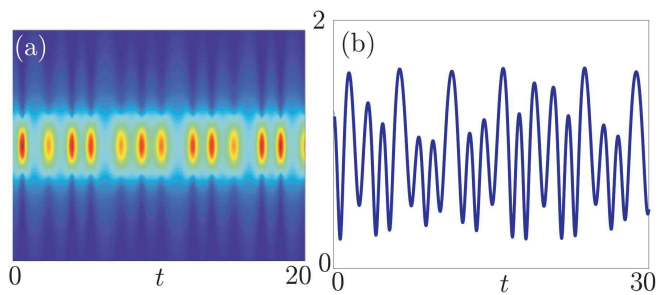,width=8.6cm}
 \caption{[Color online] Solution of Eq. (\ref{eq-physical}), given by Eq. (\ref{resonant}), for $n=1$, and $\omega^2(t)=1+\varepsilon\cos(\sqrt{2}t)$, with $\varepsilon=0.5$. The initial data for Eq. (\ref{EP}) are $\nu(0)=2^{3/4}$ and $\nu_{t}(0)=0$.
(a) Pseudocolor plot of $|\psi(x,t)|^2$  (b)  Width $\nu(t)$. 
 \label{cuasiperiodica}}
\end{figure}

We have also studied the stability of these multisoliton solutions given by Eq. (\ref{resonant}) by computing  numerically their evolution under finite amplitude perturbations. We have found that $n=1$, which corresponds to the ground state, is a stable solution while higher solitons are unstable (small perturbations take the solution far from its original profile).

\emph{Breathing solitons.-} When $\varepsilon=0, \omega(t)=1$, and then
\begin{equation}
\label{gamabreather}
\gamma(t)= \sqrt{2}\left[1+3 \cos^{2}(2t)\right]^{-1/2}
\end{equation}
 is periodic. Then, the solutions given by Eq. (\ref{resonant}) are explicit \emph{multisoliton breathing solutions}. In Fig. \ref{soluciones}, we plot some of them corresponding to $n=1,2$.  
 Only for $n=1$ (the ground state) the breathing soliton is stable. 

\emph{Quasiperiodic solutions.-} In order  to give an example of a  quasiperiodic solution of Eq. (\ref{eq-physical})  we choose $\omega_0 = \sqrt{2}$ in Eq. (\ref{omega}) ensuring that  the solutions $y_{1,2}(t)$ of the Mathieu equation (\ref{Mathieu}) belong to the stability region. Then   Eq. (\ref{omega}) has two incommensurable frequencies. In this way, $\gamma(t)$, is a quasiperiodic solution and the solutions (\ref{resonant}) show a quasiperiodic behaviour. An example of this behavior is shown in Fig.  \ref{cuasiperiodica}. 
Using Eqs. (\ref{resonant}), we can also construct \emph{multisoliton quasiperiodic solutions}.

\begin{figure}
 \epsfig{file=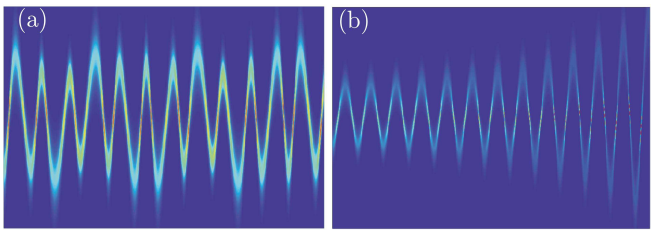,width=8.6cm}
 \caption{[Color online]  Pseudocolor plots of solutions of Eq. (\ref{eq-physical}), when the centre of mass of the soliton moves with non-zero velocity.  (a) Quasiperiodic solution, with $x\in[-3, 3]$, $t\in[0, 40]$, for $\varepsilon=0.6$ and $\nu(0)=(2\sqrt{2})^{1/2}, \nu_{t}(0)=0$. (b) Resonant increasing solution, with $x\in[-6, 6]$, $t\in[0, 40]$ for $\varepsilon=0.3$ and $\nu(0)=\sqrt{2}, \nu_{t}(0)=0$. 
  \label{solcomplejas}}
\end{figure}

\emph{Moving solitons.-} As a final example of the many possible solutions which can be constructed using this method, we will present solutions of Eq. (\ref{eq-physical}) when the centre of mass of the soliton moves with non-zero velocity. To do so, we set $f(t)$, in Eq. (\ref{f2}) equal to zero, with $\delta\neq 0$ which leads to the following equation for $\delta(t)$,
\begin{subequations}\label{cet}
\begin{equation}\label{cet1}
\delta_{tt}-2\left(\gamma_{t}/\gamma\right)\delta_{t}+4\gamma^{4}\delta=0.
\end{equation}
In this situation, the potential $v$ and the nonlinearity $g$ are still given by Eq. (\ref{potandnon}) by just taking  
\begin{equation}\label{cet2}
\alpha_{t}=\gamma^{2}(1+\delta^{2})-\delta_{t}^{2}/4\gamma^{2}.
\end{equation}
\end{subequations}
Fortunatelly, Eqs. \eqref{cet} can be solved to get 
\begin{equation}\label{solcet}
\delta(t)=\delta_{*}\cos(2\tau(t)), \quad \alpha(t)=\tau(t)+\tfrac{1}{4}\delta^2_{*}\sin(4\tau(t)), 
\end{equation}
where $\tau(t)=\int\gamma^{2}(t)dt$, and $\delta_{*}$ is an arbitrary constant.
 Hence, we can again obtain breathers, resonant solutions and quasiperiodic solutions while, at the same time the center of mass of the soliton moves in a complex way according to Eq. (\ref{solcet}). 
As examples, in Fig. \ref{solcomplejas}, we show quasiperiodic  [Fig. \ref{solcomplejas}(a)] and resonant solitons [Fig. \ref{solcomplejas}(b)] while the center of mass moves according to Eq. (\ref{solcet}).  

\emph{Conclusions.-} We have constructed explicit solutions of
 nonlinear Schr\"odinger equations in complicated yet realistic scenarios with spatially inhomogeneous and  time dependent potentials and nonlinearities. Although the range of nonlinearities and potentials for which this can be done is very wide, we have focused our study in modulated harmonic potentials and localized nonlinearities. We have explicitly calculated breathers, resonant solitons, quasiperiodic solitons and solutions with nontrivial motion of the center of mass, related to different situations of physical interest in the field of Bose-Einstein condensates. The ideas contained in this paper can also be applied to study multicomponent systems, nonlinear optical systems, higher-dimensional situations, etc., and may help in the design of potentials and nonlinearities to control the dynamics of Bose-Einstein condensates. 
 
\acknowledgements

 This work has been partially supported by grants FIS2006-04190 (Ministerio de Educaci\'on y Ciencia, Spain), PCI08-0093 (Junta de Comunidades de Castilla-La Mancha, Spain),  and POCI/FIS/56237/2004 (European Program FEDER and FCT, Portugal).


\begin{thebibliography}{10}
\bibitem{com} The relations $a_s\ll a_\bot\ll\lambda \ll a_{\|}$, where $a_s$ is the scattering length,  $a_\bot, a_{\|}$, are the transverse and longitudinal trap lengths, and $\lambda$ is a characteristic spatial scale of the wave packet, define the region in which the one dimensional approach describes correctly the physics involved.
\end{thebibliography}
  \end{document}